\documentclass[showpacs,aps,prb,twocolumn,groupedaddress]{revtex4-2}
\usepackage{graphicx}
\usepackage{textcomp}

\begin{document}


\title{Thickness dependent rare earth segregation in magnetron deposited NdCo$_{4.6}$ thin films studied by Xray reflectivity and Hard Xray photoemission}


\author{J. D\'iaz,$^{1,2}$}
\email[]{jidiaz@uniovi.es}
\author{J. Rodr\'iguez-Fern\'andez,$^{1,2}$}
\author{J. Rubio-Zuazo,$^{3,4}$}
\affiliation{$^{1}$Universidad de Oviedo, Calle Leopoldo Calvo Sotelo, 18, Oviedo 33007, Spain}
\affiliation{$^{2}$CINN (CSIC-Universidad de Oviedo), 33940 El Entrego, Spain}
\affiliation{$^{3}$SpLine Spanish CRG Beamline at the ESRF, ESRF-BP 220-38043 Grenoble cedex, France} 
\affiliation{$^{4}$Instituto de Ciencia de Materiales de Madrid-ICMM/CSIC Cantoblanco E-28049 Madrid, Spain}
\affiliation{}


\date{\today}

\begin{abstract}
Magnetic anisotropy in disordered rare-earth-transition metals (RE-TM) compounds arises from RE atoms occupying asymmetric environments within the TM lattice. However, the underlying mechanism that promotes such environments remains not fully understood. In this study, we investigate amorphous NdCo$_{4.6}$ thin films deposited by magnetron sputtering, where the magnetic anisotropy evolves with thickness from in-plane to out-of-plane orientation above 40 nm. X-ray reflectivity measurements revealed the progressive formation of an additional layer between the 3 nm Si capping layer and the NdCo film with increasing film thickness. To probe the composition and distribution of Co and Nd near the surface, Hard X-ray Photoemission Spectroscopy (HAXPES) was performed on films ranging from 5 nm to 65 nm in thickness using incident photon energies of 7, 10, and 13 keV. These correspond to inelastic electron mean free paths of 7.2-12.3 nm in cobalt. The cobalt atomic concentration, deduced from HAXPES at the Nd 3d and Co 2p excitations, was consistently below the nominal value and varied with both thickness and photon energy. This indicates segregation of RE atoms at the film surface, which becomes more pronounced with increasing thickness. Background analysis of the Co 2p and Nd 3d peaks supports this conclusion. The thickness of the Nd segregated layer is estimated in 2-3 nm. Nd segregation provides evidence for Nd diffusion toward the surface during thin film growth, what demonstrates that incorporating neodymium into the cobalt lattice incurs an energy cost which is associated with strain due to the volume mismatch between the two elements. Reducing this strain energy promotes anisotropic atomic environments for those Nd trapped into the cobalt lattice, thereby explaining the emergence of perpendicular anisotropy and its dependence on film thickness in these RE-TM compounds.

 \end{abstract}


\maketitle

\section{INTRODUCTION} 
Amorphous NdCo$_{4.6}$ is a ferromagnetic intermetallic RE-TM compound with modest perpendicular magnetic anisotropy (PMA) at room temperature (RT), whose anisotropy energy quality factor $Q=K_{u}/4\pi M_{s}$ is less than 1, being $K_{u}$ the PMA energy and $M_{s}$ the total magnetization of the alloy. This property makes it less anisotropy dominant and more flexible for implementation in various applications. They are produced at RT and can be stacked or combined with other materials giving rise to devices with interesting functionalities. Most of them are based on striped magnetic domains formed in thin films which have interesting properties such as a well-defined constant period, reconfigurable orientation, rotatable anisotropy \cite{rotatable_anisotropy}, and topological textured features\cite{AureNat_2020}. This compound is used, for instance, in bidimensional array composites with tunable exchange bias\cite{NdCo_SoftHard}, in reconfigurable magnonic devices \cite{Luis_APL,Polaco_SpinWaves} and in domain wall racetracks \cite{Victoria_movingwalls}.

The magnetic moment of a RE element originates from its strongly spin-orbit-coupled 4f orbital, which is largely isolated from the valence electrons. Magnetic coupling between the TM spin and the 4f electrons of the RE is facilitated by intratomic exchange interactions involving the 6s and 5d valence electrons, which are themselves spin-polarized by the 3d electrons of the TM \cite{Intratomic_RE_STM}. This coupling is relatively weak, on the order of meV, and competes with the crystal field at the RE site, which determines the orientation of the 4f angular momentum \cite{buschow_RMreview77,Intratomic_RE_STM,RE_anisotropy}. This orientation defines the anisotropy of the Nd magnetic moments at each of their locations in the compound. The PMA observed in RE-TM compounds emerges when there is a preference in the symmetry axis of the RE crystal fields for the out-of-plane alignment of its magnetic moment. Understanding the mechanism that originates such a preferential asymmetry in the atomic environment of the RE within the alloy is therefore key for controlling the PMA properties of RE-TM alloys \cite{Pair_ordering,amorphous_anelastic_diffraction,TbFeCo_anelastic_diffraction}. Structural asymmetry has been confirmed in other amorphous RE-TM systems, such as amorphous TbFe, through techniques including diffraction \cite{TbFeCo_anelastic_diffraction}, transmission electron microscopy \cite{TbFe_TEM}, and extended X-ray absorption fine spectroscopy (EXAFS) \cite{TbFe_EXAFS,NdCo_EXAFS}.  

Understanding how RE enters this type of preferential asymmetric environment is an old problem owing to the complexity of the system\cite{Boltzman_anisotropy_model,Selective_resputtering,Pseudocrystalline_model}. However, the most accepted explanation for thin films deposited using magnetron sputtering methods, which have incident atom energies above those of thermal evaporation \cite{sputtering_model}, is based on lowering the surface energy during thin film growth\cite{anisot_model_surfaceE_b,anisot_model_surfaceE}. Following this explanation, a majority of RE are in specific asymmetric environments because the corresponding adatom locations during growth are those that reduce the surface energy the most, creating a structural texture that gives rise to the resulting anisotropic environments. Thus, to control the PMA and the magnetic properties of the NdCo compounds it is neccesary to elucidate the parameters that condition the adatom diffusion during thin film growth.

The mixing of a TM and a RE in a solid is conditioned by the large differences in their atomic volume, which is three times lager for Nd (3.4$\times 10^{-29}$ m$^{3}$) than Co (1.1$\times 10^{-29}$ m$^{3}$). This should cause strain that the system might try to avoid by segregating the larger volume atom\cite{segregation_1979,segregation_TM}. Segregation has been reported in NdCo and in DyCo alloys using Xray magnetic circular dichoism (XMCD) \cite{NdCo_PRB,DyCo_XMCD}. RE segregation has also been  reported in other different RE-TM compounds such as TbFeCo \cite{TbFeCo_segre_PRM} and GdCo \cite{GdCo_segre_PRB}. The consequent gradation in the concentration of RE at the surface has given rise to interesting magnetic effects in ferrimagnetic DyCo compounds \cite{DyCo_FerriGrad}. Interestingly, thickness-dependent experiments performed on GdCo and TbCo ferrimagnetic compounds using metallic buffer layers have detected magnetic soft and even dead layers at the interface with the substrate\cite{TbCo_segregation,GdCo_segre_Manu}, which may indicate that the RE distribution within the RE-TM compound film could be modified by choosing the right interface.

Previous XMCD measurements of the studied NdCo films performed at temperatures ranged from 10 K up to RT demonstrated the out-of-plane magnetic anisotropy of Nd and the in-plane anisotropy of Co, identifying Nd as the origin of the PMA in the alloy \cite{NdCo_PRB}, and consequently proving that a majority of Nd atoms have to be in asymmetric atomic environments. EXAFS spectroscopy performed in 100 nm thick samples was not able to identify them due to a vanish EXAFS signal for neodymium\cite{NdCo_EXAFS}, likely due to disorder \cite{EXAFS_limits} and an apparent low Nd concentration in the bulk. However, the measurements evidenced a highly compressed cobalt sublattice. Cobalt atoms presented a short-range order equivalent to a fcc lattice with a first neighbor distance of 2.40 \AA. This is equivalent to a reduction in volume of 11$\%$ with respect to fcc cobalt at ambient pressure \cite{Compressed_Co}. This bond length reduction is not present in pure cobalt films deposited in the same magnetron sputtering system or in other amorphous alloys like Co-Si at similar concentrations \cite{CoSi_Co_EXAFS}, indicating that introducing Nd into the cobalt lattice causes a significant compressive strain. This short-range ordered cobalt environment was not comparable to the expected in crystalline NdCo$_{5}$ which have 6 first nearest neighbors at 2.45 \AA. It also indicates larger than expected distances between Nd atoms within the bulk of the alloy. The measurements done at 10K, where thermal expansion of the film and the substrate is reduced with respect to RT, reported an anomalous expansion of the cobalt lattice, what confirmed its compressive strained structure.

The present experiment shows that the PMA properties of our NdCo$_{4.6}$ are thickness dependent. PMA appears above a threshold thickness between 30 to 40 nm and gradually increases until it reaches its saturation value at a thickness below 100 nm, indicating that it is built during the growth process of the film, what is coherent with an increasing compressive strain as thickness increases. In order to obtain information on the possible diffusion process occurring during thin film growth that drive the system to its PMA properties, X-ray reflectivity (XRR) and HAXPES\cite{HAXPES_rev1,HAXPES_rev1} were applied to different thin-film thicknesses. Both techniques are not invasive and they are sensitive to the bulk and the interfaces of the deposited films. XRR is highly sensitive to layer interfaces and should detect changes in thickness, electron density and roughness. However, it cannot provide the precise nature of the elemental composition of the layers and interfaces without ambiguity. The electron densities of cobalt and neodymium may not be large enough to be distinguished if there is some intermixing. Photoemission spectroscopy is an elemental-sensitive probe with probe depths that can be about three times the inelastic electron mean free path ($\lambda_{IMFP}$) of the photoemitted electrons\cite{e_IMFP}. By using hard Xrays, HAXPES reaches larger probing depths than Xray absorption spectroscopy, when using secondary electron detection \cite{Probing_depth_XAS}, or Xray photoemission (XPS). Moreover, the probe depth can be modified by the incident photon energy and/or by choosing the specific excitation peaks\cite{HAXPES_rev1}. This allows the study of the composition and chemistry of the top layer region of the samples underneath the 3 nm thick silicon capping layer and checks the extension of the possible segregated layers by changing the probe depth.


The experiment is especially centered in the information collected by HAXPES to determine the gradation in the composition of thin films\cite{HAXPES_Juan_perfil}. This information comes from the measured atomic relative concentrations of cobalt to neodymium,  the valence band spectra, the shape of the photoemitted Co 2p and Nd 3d peaks and the shape and intensity of their background due to inelastically scattered electrons\cite{Tougaard_good,Tougaard_visual}. The results of the experiment are consistent with an effective segregated layer of neodymium in all the thin films analyzed whose effective thickness is of the order of 2 to 3 nm, demonstrating that there is a diffusion of neodymium atoms toward the surface. This is consequent with the compressive strain caused by the incorporation of neodymium in cobalt. This also indicates that Nd atoms trapped within the cobalt lattice must adopt atomic environments that, in average, should reduce the compressive strain energy of the cobalt lattice. This means that they should be more relaxed in the vertical direction where lattice expansion is energetically favorable, explaining the origin of their PMA properties.

\section{EXPERIMENT} 
The samples were co-deposited using three confocal magnetron sputtering guns, Co, Nd and Si. The Co gun was normal to the substrate whereas the Nd and Si guns formed an angle of 30$^{\circ}$ on opposite sides to the normal of the sample. The Ar pressure during deposition was 3$\times10^{-3}$ mbar, with a base pressure of 1$\times10^{-8}$ mbar. The thin films were deposited on the native oxide of boron doped silicon substrates at a rate of 2 \AA$ /$sec. The voltage bias and power applied to the magnetron guns during deposition were approximately 430 V and 0.10 kW for Co, and 330 V and 0.03 kW for Nd. All the films had a protective capping layer of 3 nm thick silicon.  Deposition rates and concentration were calibrated using a quartz balance and XRR. 

The atomic concentration of the NdCo compound, NdCo$_{4.6}$, was chosen because it gives the highest magnetic moment and PMA energy $K_{u}$. Pure cobalt thin films of the same thickness were deposited in the same set of experiments to obtain a reference for cobalt magnetizations and layer interfaces. The magnetization and hysteresis loops (HL) of the samples were measured using a vibrating sample magnetometer (VSM) and an alternating gradient magnetometer (AGM).  XRR measurements were performed in a two-axis difractometer (PANalytical X'pert) adapted to reflectometry measurements using a photon energy of 8.04 keV (Cu $K_{\alpha^{*}}$). 

HAXPES was performed at the ESRF synchrotron in the BM25 beamline \cite{spline}. Measurements were done at UHV and RT using a cylindrical analyzer\cite{a_cilindrico} with a pass energy of 100 eV using incident photon energies of 7, 10 and 13 keV corresponding to a $\lambda_{IMFP}$ for cobalt of 7.2 nm, 9.8 nm and 12.3 nm respectively. The angle of incidence of the X ray beam was 5$^{\circ}$ with respect to the sample plane. The electron analyzer collected the photoemitted electrons near the orthogonal direction from the incoming beam, at 15$^{\circ}$ with respect to the normal direction of the sample. 

\section{RESULTS and discussion} 

\subsection{Magnetic Characterization} 

\begin{figure}
\includegraphics[width=8 cm]{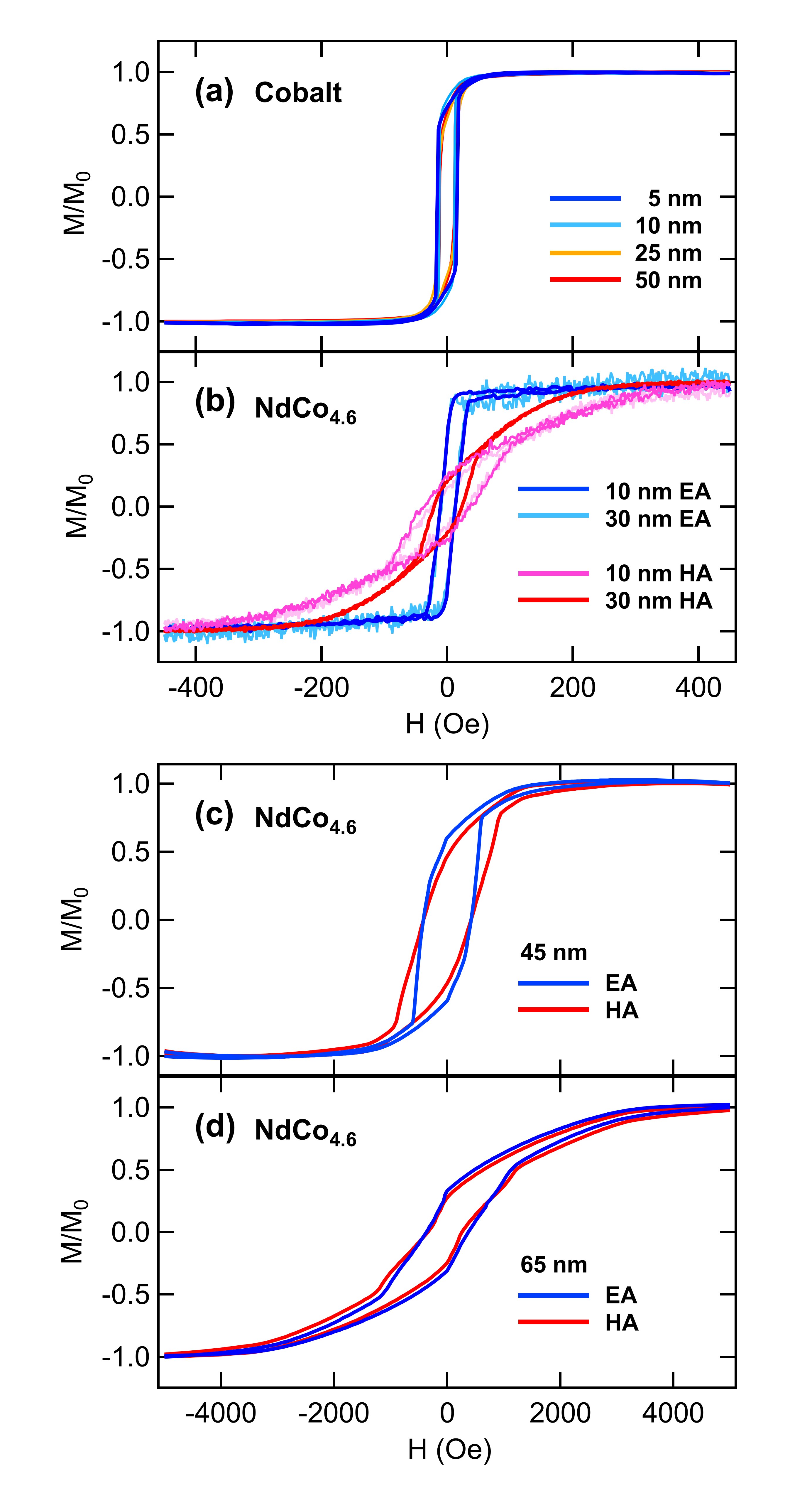}
\caption{ Hysteresis loops measured by VSM of the in-plane magnetization of (a) pure cobalt of thickness 5, 10, 25 and 50 nm, and NdCo$_{4.6}$ thin films along the direction parallel (blue) and perpendicular (red) to their in-plane EA of thickness (b) 10 nm and 30 nm, (c) 45 nm, (d) 65 nm.\label{fig_ciclos}}
\end{figure}

Figure \ref{fig_ciclos} shows the in-plane field dependent magnetization (HLs) for pure cobalt and NdCo$_{4.6}$ thin films of similar thickness obtained by magnetometry at RT. The shape of the HLs of cobalt was independent of the film thickness. This is shown in figure \ref{fig_ciclos}(a) where the HLs of the cobalt thin films with thicknesses ranging from 5 to 50 nm are superimposed. Their coercive field $H_{c}$ is  20 Oe, which is practically the same as that measured in the NdCo$_{4.6}$ thinner films (fig. \ref{fig_ciclos}(b)). Owing to the normal incidence of Co atoms during thin film deposition, they have negligible in-plane magnetic anisotropy. Their magnetic remanence $M_{H=0}/M_{s}$ was close to 60$\%$. 

The HLs measured in the NdCo$_{4.6}$ samples with thicknesses below 40 nm are typical of soft magnets with in-plane anisotropy. Figure \ref{fig_ciclos}(b) shows them superimposed because they barely change with thickness. Their in-plane magnetic easy axis (EA) is perpendicular to the plane of incidence of the Nd and Si atoms, confirming the oblique incidence origin of their in-plane anisotropy\cite{Oblique_Carlos_2014,Oblique_FeCoB_2020,Oblique_STM_2006}. Their EA HLs have almost 100$\%$ remanence, indicating that PMA energy is small compared with the in-plane anisotropy at these thicknesses. The hard axis (HA) has a saturation field, $H_{s}$, of approximately 200 Oe at 30 nm, increasing to almost 400 Oe at lower thicknesses. The opening of the hysteresis loop in the HA might indicate a possible out of plane anisotropy component. As the thickness of the samples increased above 40 nm, the $H_{c}$ increased, the remanence decreased and the in-plane $H_{s}$ increased, as shown in figure \ref{fig_ciclos}(c). These are all signs of increasing magnetic anisotropy in the alloy. The HL of the 65 nm thick sample, in figure \ref{fig_ciclos}(d), shows a $H_{s}$ which is 10 times the measured in the samples with thickness lower than 40 nm. In addition, there is almost no difference between the HL measured along the in-plane EA and the HA: the in-plane anisotropy becomes overwhelmed by the out-of-plane anisotropy. 

The magnetization of both cobalt and NdCo$_{4.6}$ thin films is approximately 18$\%$ lower at 5 nm, but reaches its saturation value for thicknesses above 20 nm, indicating cobalt interdiffusion at the silicon interfaces. XRR measurements shows interfaces widths of the order of 4 to 5 \AA. The dominant in-plane magnetic anisotropy at low thickness, where the magnetization is the lowest, proofs that its PMA energy is smaller than at higher thickness. The magnetization of the cobalt thicker films was the expected for an atomic magnetic moment of Co of 1.7 $\mu_{B}$. The magnetization of the  thickest NdCo films was approximately 800 emu/cm$^{3}$. The expected magnetization of the NdCo$_{4.6}$ compound, $M_{s}$, can be calculated from the atomic concentration, the neodymium and cobalt atomic volumes determined from their related densities (8.9 gr/cm$_{3}$ and 6.9 gr/cm$_{3}$), and their magnetic moments $m_{Co}$ and $m_{Nd}$. For the atomic concentration of NdCo$_{4.6}$, $M_{s}(emu)=501\cdot m_{Co}(\mu_{B})+110\cdot m_{Nd}(\mu_{B})$. The expected magnetic moment of the 4f orbital in Nd is $gJ=3.27 \mu_{B}$ related to the electronic configuration $[Xe] 4f^{3}6s^{2}5d^{1}$. However, this value is difficult to reach at RT because of the weak exchange coupling between Co and Nd atoms, disorder and possible presence of Nd unbonded to Co \cite{NdCo_PRB,DyCo_XMCD}. In addition, cobalt reduces its magnetic moment not only by bonding to neodymium but because of the reduced cobalt bond length registered by EXAFS \cite{NdCo_EXAFS,mCo_vs_pressure}, which contributes to the reduction of the atomic magnetic moment of Nd. This was observed in previous XMCD measurements\cite{NdCo_PRB} in samples with similar atomic concentrations. By taking into account all these factors, it is estimated that the  magnetic moment of Co and Nd in these samples is roughly 1.35 $\mu_{B}$ and 1.1 $\mu_{B}$, respectively. As it will be shown from the HAXPES data, these relatively low magnetic moments can hardly be attributed to oxygen contamination. 

The anisotropy energy of the analyzed samples can be estimated from $K_{u}=\frac{1}{2}H_{s}M_{s}$. PMA energy ($K_{u}$) is 0.84$\times 10^{6}$ erg/cm$^{3}$ in the 45 nm film and 2.4$\times 10^{6}$ erg/cm$^{3}$ in the 65 nm thick film. The in-plane anisotropy energy of the samples with 30 nm thickness or lower is approximately 0.3$\times 10^{6}$ erg/cm$^{3}$. The apparent sharp transition from in-plane to out-of-plane anisotropy, which happens in a thickness difference of only 10 nm, might be explained because the competition between the shape anisotropy, which forces an in-plane magnetization, and the PMA. As the thickness increases, a gradual increase in PMA energy might occur. However, because of the shape anisotropy, the magnetic film needs to be arranged in a stripe magnetic domain configuration. This happens at a critical thickness, which is the size of the stripe domains. This size depends on $K_{u}$ and the magnetic exchange in the alloy\cite{Hubert}. After this transition, figure \ref{fig_ciclos} shows that the PMA energy increases gradually with thickness. The analysis of thicker samples shows that the saturation value for the PMA energy is reached around 65 nm.

\subsection{Xray Reflectivity} 
\begin{figure}
\includegraphics[width=8 cm]{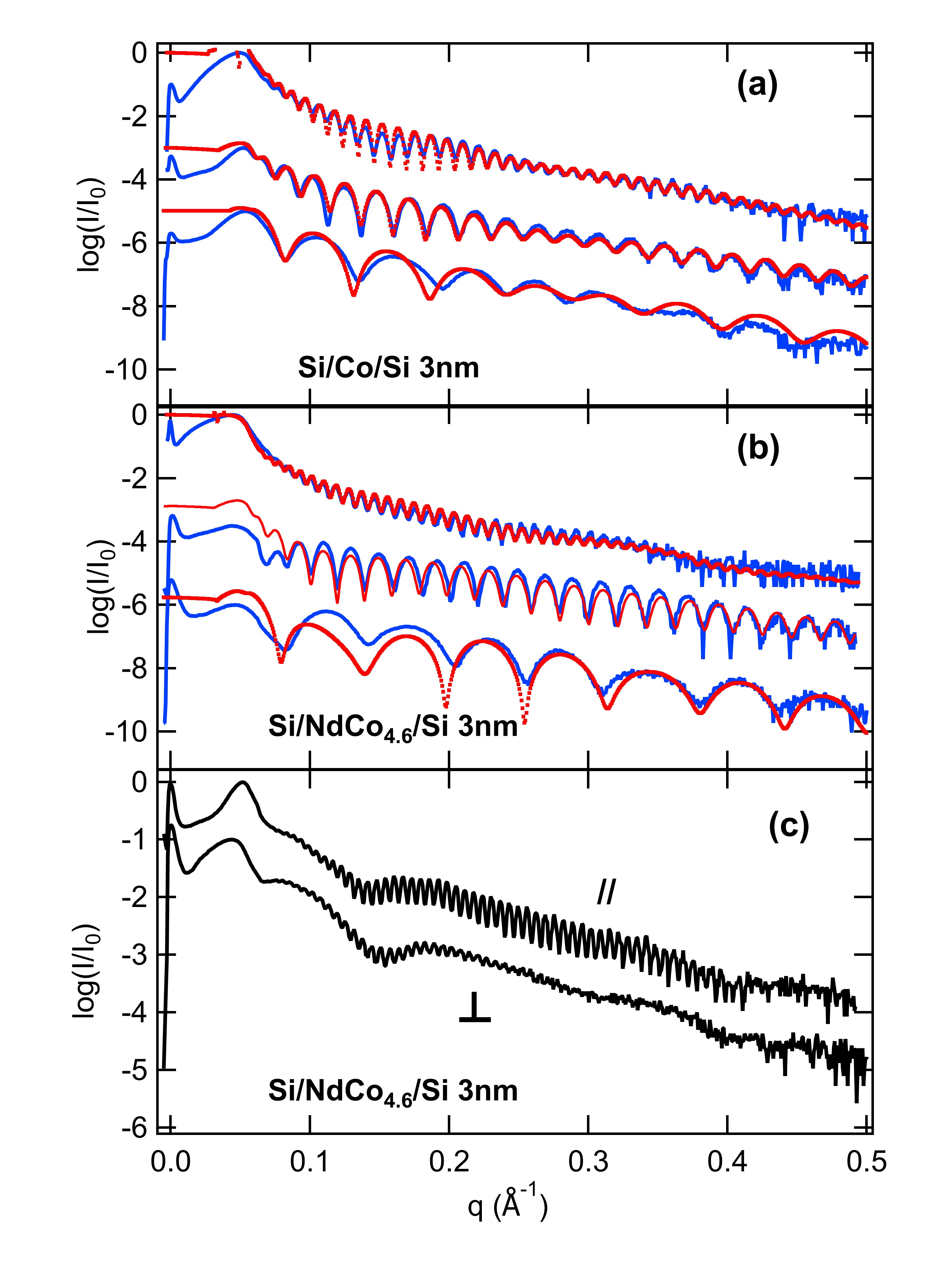}
\caption{ XRR curves (blue) and their fits (red) of (a) pure cobalt thin films of thickness, from top to bottom, of 50, 30 and 10 nm; (b) NdCo$_{4.6}$ thin films of 65, 30 and 10 nm thickness. (c) XRR curves of 95 nm thick  NdCo$_{4.6}$ alloy obtained with the incident wave vector $k_{i}$  parallel ($//$) and perpendicular ($\bot$) to the plane of incidence of the neodymium and silicon magnetron beams \label{fig_Xrflx}}
\end{figure}

Figure \ref{fig_Xrflx} compares the XRR measurements of the NdCo$_{4.6}$ analyzed samples, except for the 5 nm thick sample, with the corresponding pure cobalt reference samples. The overlapping curves in red represent the related fitting curves. They have larger misfits in the thinner samples indicating a higher incertitude for the parameters obtained for those films. The fitting of the reflectivity curves were done by calculating the Xray interferences between the reflections at the different layer interfaces using the Abeles formalism\cite{Gibaud2009}. The parameters to adjust for each layer included in the simulation were its thickness, complex refractive index (which is related to the electron density of the material) and roughness/diffusion of the interface with the adjacent top layer. A first guess of the thickness components present in the curves was obtained from their Fourier transform. The high frequency oscillations (Kiessig oscillations) in the reflectivity curves are mainly related to the thickness of the thicker layer. The low frequency modulation in the amplitude of these oscillations corresponds to thinner layers which in this case are the silicon capping layer plus any additional layer related to segregated material. This low-frequency modulation is different in NdCo$_{4.6}$ thin films compared to that of pure cobalt. 
The best fits for the reflectivity curves of the cobalt films consist of a single layer of cobalt plus a capping layer of approximately 1.1 nm thickness with an electron density slightly higher than that of silicon. The exception to this fit is the 50 nm thick thin film which requires a top layer of 2.2 nm, which is slightly less dense than that of silicon, and a 1.2 nm thick interdiffusion layer. 

NdCo$_{4.6}$ thin films with thicknesses below 40 nm require a single layer on top of the alloy, as in pure Co thin films. However, its thickness is larger (3 nm). The surface roughness increased as the thickness increased. For thicknesses greater than 40 nm, the simulations required two top layers. The surface of the first of these layers is the one in contact to air. It is rougher and thinner (of about 1.5 nm) than the layer underneath, which is about 4 nm thick and it has a higher electron density. The roughness of the interface between these two layers  is of only 3 to 4 \AA. The interface with NdCo$_{4.6}$ layer is rougher, of about 6 \AA, indicating some interdiffusion between the layers.

The second layer in addition to the capping top layer might be associated with segregated neodymium because this layer does not appear in the pure cobalt films. However, the electron density obtained from the best reflectivity simulations was slightly higher than half of the expected value for pure Nd. Attempts to  fit the curves using  metal neodymium density yielded worse fits and thinner layers. This additional layer should then be oxidized or silicized neodymium to match the observed electron densities. A better identification of the elements segregated at the top layers is done by HAXPES. 

Figure \ref{fig_Xrflx}(c) presents two reflectivity curves of a 95 nm thick NdCo$_{4.6}$ film, where the low-frequency amplitude modulation associated with the top double layer is more pronounced than in the 65 nm thick layer shown in (b). The fitting parameters for these top layers are essentially identical. Both reflectivity curves were measured on the same sample, with the X-ray wave vector $\textbf{k}_{i}$ oriented either parallel ($\|$ sign) or perpendicular ($\bot$ sign) to the plane of incidence of the Nd and Si magnetron guns.

The differences in the amplitude of the Kiessig oscillations between the two orientations arise from variations in the roughness at the substrate interface. This asymmetry is also observed in thinner films, though the effect is less significant. It originates directly from the oblique incidence of Nd atoms during thin-film growth, producing a shadow effect which establishes a preferential diffusion direction for Co perpendicular to the Nd incidence plane. As a result, the Co deposited tend to expand on the surface in the direction perpendicular to the plane of incidence creating a corrugation in the roughness of the interfaces which is anisotropic.

This mechanism explains the in-plane magnetic anisotropy in these films as a form of shape anisotropy \cite{Oblique_STM_2006}. This anisotropy might be also driven by the reduced magnetization and weaker magnetic exchange of Nd with Co compared to Co–Co interactions at RT if the distribution of neodymium is not completely homogeneous, i.e., if there is an in-plane asymmetry for the Nd environments \cite{FeSi_EXAFS}. However, the large energy difference between the in-plane and out-of-plane anisotropies, of one order of magnitude even in thicker samples where shadowing effects are more evident, indicates that the perpendicular magnetic anisotropy (PMA) is of a different nature than the in-plane one. Also, it cannot be attributed to effects like columnar shape anisotropy as occurs in other TM amorphous alloys with PMA  \cite{PMA_amorphous}. Instead, the Nd environments responsible for PMA are those where neodymium atoms are more strongly exchange-coupled to cobalt. This interpretation is consistent with the observed increase in magnetization and PMA energy as the films are cooled to lower temperatures \cite{NdCo_PRB,NdCo_EXAFS,DyCo_XMCD}.

\subsection{HAXPES} 
\subsubsection{Valence band region}
\begin{figure}
\includegraphics[width=8 cm]{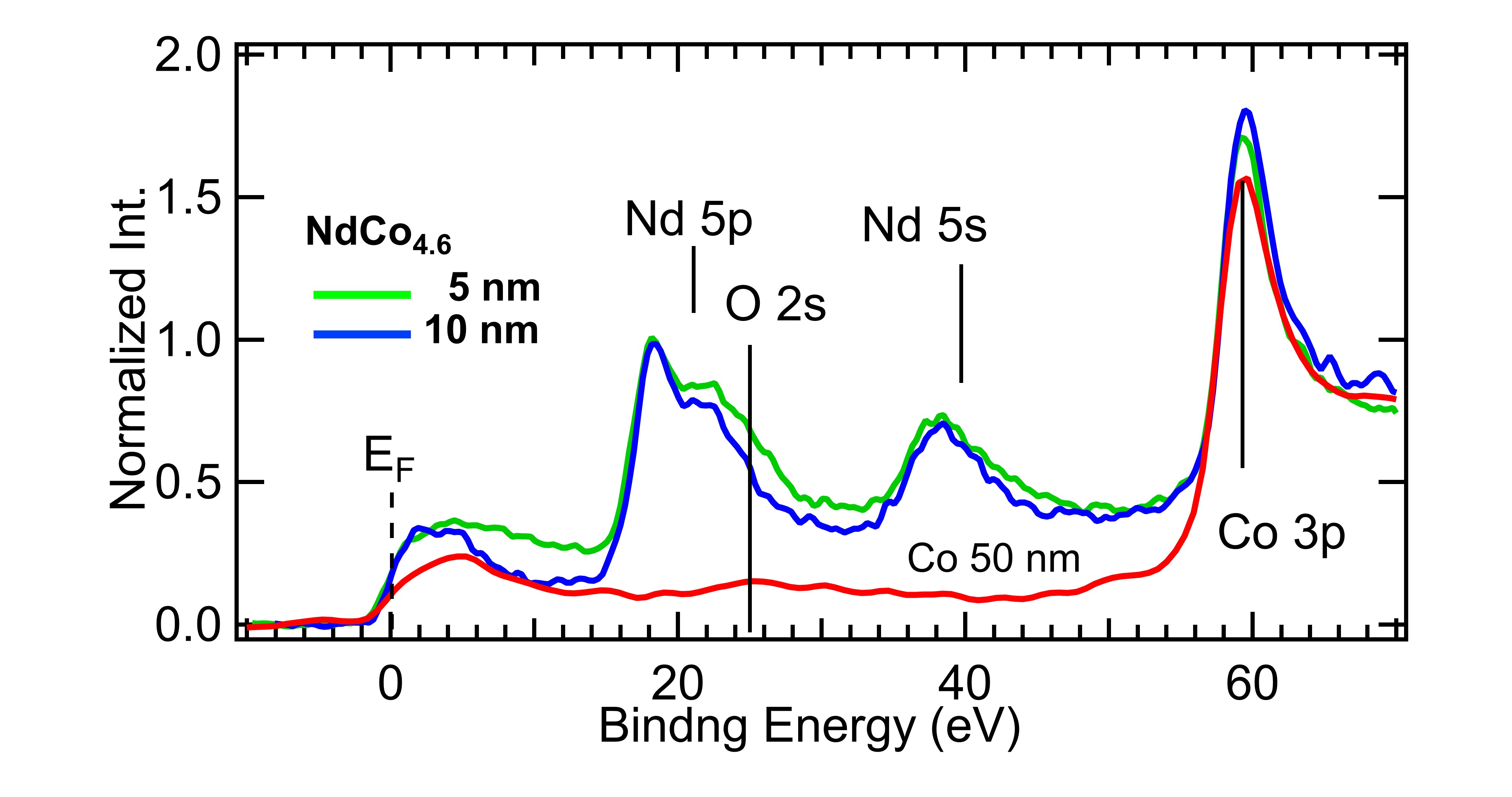}
\caption{Comparison of the HAXPES spectra obtained at 7 keV photon energy of the VB region in the pure cobalt 50 nm thick film (red) and in the 5 nm (green) and 10 nm (blue) NdCo$_{4.6}$ films.\label{fig_vb}}
\end{figure}

Figure \ref{fig_vb} compares the region near the valence band (VB) of NdCo$_{4.6}$ films at two different thickness, 5 nm and 10 nm, with the VB of a pure cobalt thin film obtained with 7 keV Xray photons. The spectra were normalized to the intensity of the Co 3p peak at 59.5 eV. The spectrum of the 5 nm thick alloy has a signal related to silicon and oxygen excitations in the region between 6 eV ang 16 eV, corresponding to the VB of SiO \cite{VB_SiO}. This film has a much larger contribution from the silicon substrate than the other analyzed films because its thickness was lower than the $\lambda_{IMFP}$ of the photoemitted electrons for the 7 keV incident photon energy. The VB spectrum of the 10 nm thick alloy is representative of the NdCo$_{4.6}$ thicker films because it is no different from the obtained in the 30 and 65 nm thin films. Therefore, the differences between the shown VB spectra indicates the regions where Si and O excitations are expected. The difference in shape between the Nd 5p peaks in the spectra of the NdCo$_{4.6}$ compound is likely caused by the O 2s peak located at a binding energy (BE) of approximately 25 eV, whose more pronounced intensity comes from the native SiO interface of the substrate. The intensity of this O 2s peak is almost undetectable in the VB spectrum of the pure cobalt film. 

The intensity in the region near the Fermi level is mainly related to the 3d electrons of Co and to the 4f, 6s and 5d electron states of Nd, with some contribution from the Si states of the capping layer. The differences between the VB of pure cobalt and the RE-TM compound indicates that the most visible occupied states in the compound thin films should be those of Nd \cite{Nd_peakfit}. This is unexpected from the nominal atomic RE concentration in the alloy which is 4.6 times smaller than that of Co. The photo ionization cross sections for Nd and Co states in this range of binding energies are similar and, therefore, it indicates that neodymium atoms should be at the fore front of the HAXPES probed layer. 

\subsubsection{Co 2p and Nd 3d spectra}
\begin{figure}
\includegraphics[width=8 cm]{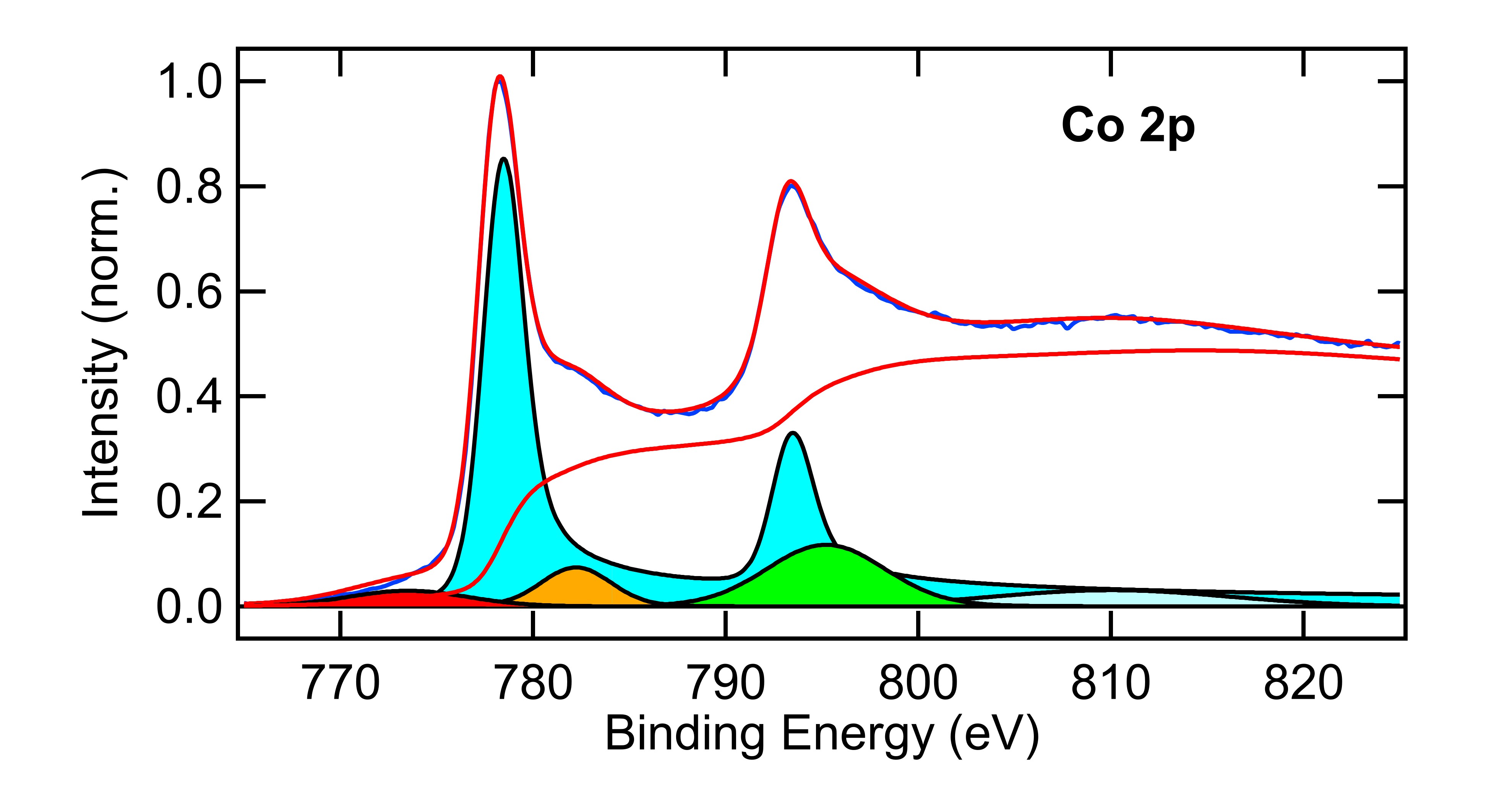}
\caption{Fitting components and Shirley type background of the Co 2p HAXPES spectrum of the 65 nm thick NdCo$_{4.6}$ film obtained at 7 keV photon energy.\label{Fig_Co_comp}}
\end{figure}

\begin{figure}
\includegraphics[width=8 cm]{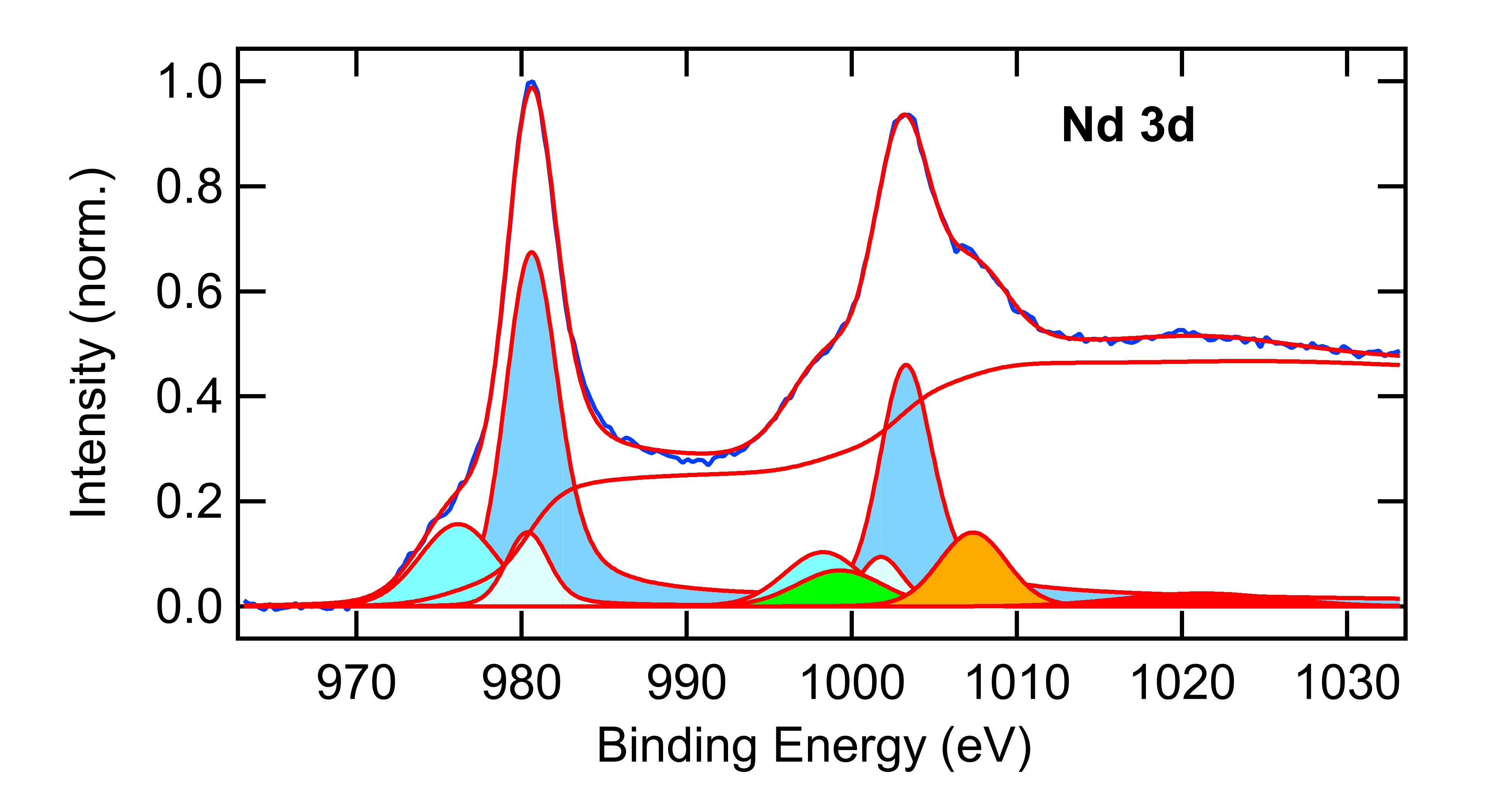}
\caption{Fitting components and Shirley type background of the Nd 3d HAXPES spectrum of the 65 nm thick NdCo$_{4.6}$ film obtained at 7 keV photon energy.\label{Fig_Nd_comp}}
\end{figure}

A more quantitative method of proving segregated Nd at the surface of thin-film alloys is to calculate the atomic concentration determined by measuring the areas of the Co 2p and Nd 3d peaks of the analyzed thin films. These measurements were performed at different probing depths by varying the energy of the x-ray photon. The components used to fit the Co 2p and Nd 3d spectra were the same for all the three probing  photon energies. The only difference was the width of the peaks due to the decreasing resolving energy of the incident beam and analyzer with increasing photon and kinetic energy. Figures \ref{Fig_Co_comp} and \ref{Fig_Nd_comp} shows the components of the spectra fits. Each component is the convolution of a Lorentzian, Gaussian and a Doniac-Sunjic function with widths related to the intrinsic excitation energy width, instrumentation plus structural disorder, and density of states at the Fermi level, respectively. The background used for the analysis was Shirley-type \cite{Shirley}. The Co 2p spectra were fitted using two main components related to the $2p_{3/2}$ and $2p_{1/2}$ spin-orbit split excitations with a splitting energy of 15 eV, similar to cobalt metal. Two additional components were located at higher BE which may be associated with cobalt oxidation. However, as will be discussed further, none of the components of the O 1s peak had the expected low BE for oxygen bonded to a metal. A third additional component, broad and of much lower intensity, was required in the fits, which was positioned at lower BE that the $2p_{3/2}$ peak, as a pre-threshold excitation. Finally, a broad peak was used to fit the bulk plasmon excitation located at approximately 810 eV BE. The position of the Co $2p_{3/2}$ peak was set at 778.4 eV in the pure cobalt films. The alignment of the pure cobalt and NdCo films spectra using the position of their Fermi level did not detect displacements from this position in any of the samples. Moreover, the fitting of the Co 2p spectra of the NdCo compounds  was similar to that of the pure cobalt thin films, which differentiated only by the width of the peaks. 

The components used to fit the Nd 3d spectrum were very similar to those used by Maiti et al.\cite{Nd_peakfit} for Nd in the intermetallic compound Nd$_{2}$PdSi$_{3}$. It consists on three spin-orbit split peaks, Nd 3d$_{5/2}$ and Nd 3d$_{3/2}$, positioned at different energies with different splitting energies, which corresponds to the different electron configurations that the Nd electronic structure adopt after the core hole is created. The most intense Nd 3d$_{5/2}$ excitation peaks at 980.6 eV and it has an splitting energy of 22.7 eV, which is the expected value for Nd in a metal. The other two Nd 3d$_{5/2}$ peaks are located at lower binding energies (976 eV and 980 eV) with spin-orbit splitting energies of 21.9 eV and 21.4, respectively. Two additional relatively broad components at  999.2 eV and 1007 eV binding energies are associated to plasmon excitations. None of these components have been clearly associated with oxidized states which are expected to have higher Nd 3d$_{5/2}$ binding energies than the observed.

\begin{figure}
\includegraphics[width=8 cm]{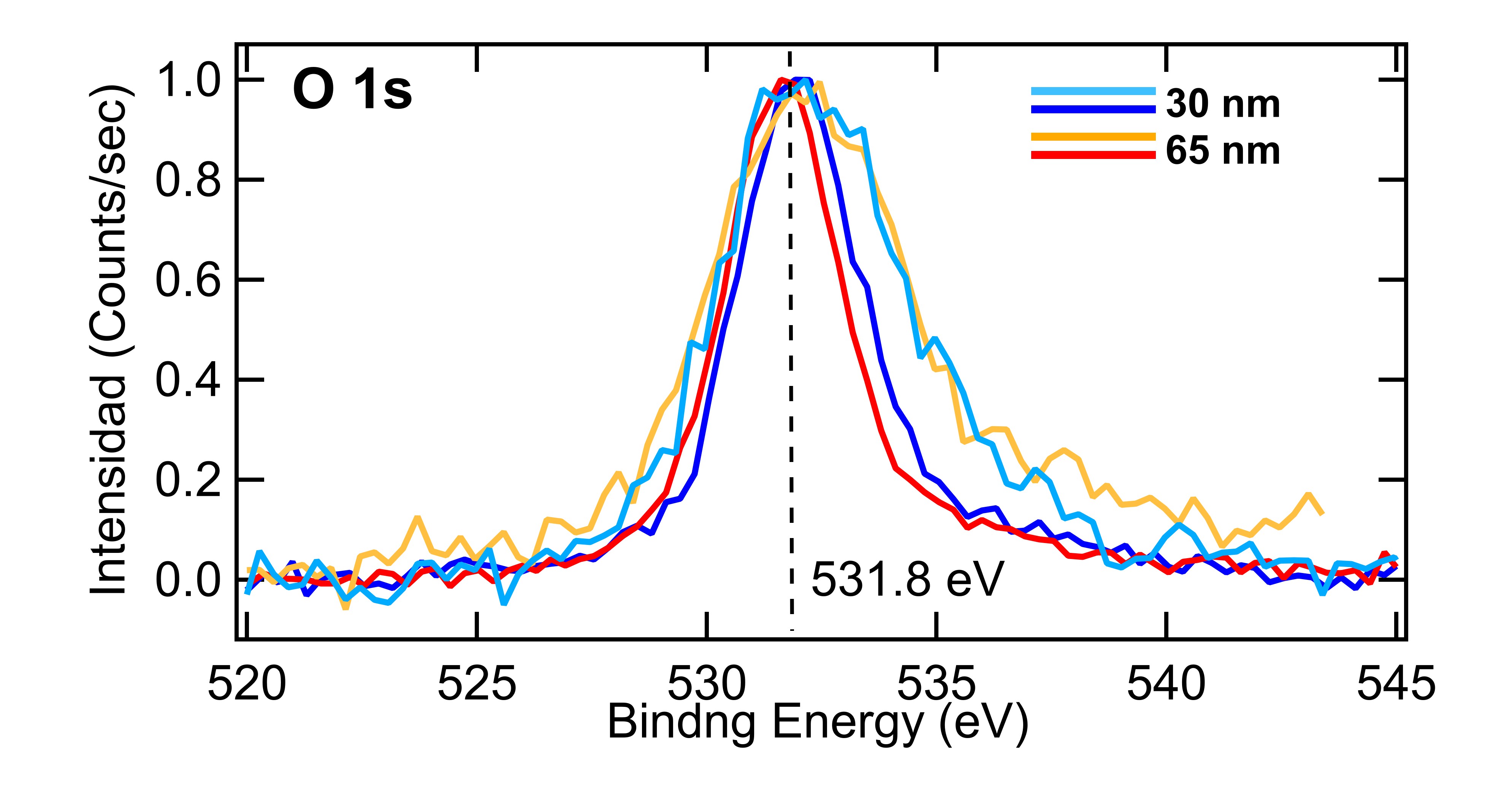}
\caption{Comparison of the O 1s HAXPES spectra taken in NdCo$_{4.6}$ films obtained at 7 keV photon energy (65 nm red, 30 nm blue) and 13 keV (65 nm orange, 30 nm light blue) \label{Fig_O1s}}
\end{figure}

Figure \ref{Fig_O1s} compares the O 1s spectra of the 30 nm and 65 nm films taken at two different photon energies, 7 keV and 13 keV. These values were normalized in terms of the intensity. The intensities of the peaks obtained at 13 keV were significantly lower than those at 7 keV. This reduction is lower than expected from the decrease in cross section of the O 1s peak, what might indicate the presence of some oxygen within the photoemission probed layer. The fit of the O 1s peaks shows components at BE above 531 eV. This BE is higher than the expected value for oxidized metals\cite{O1s_XPS}. The splitting of the Si 2s peak into two components (not shown) and the absence of an electron energy loss background step indicates that this oxygen comes mainly  from the oxidation of the silicon capping layer.

\begin{figure}
\includegraphics[width=8 cm]{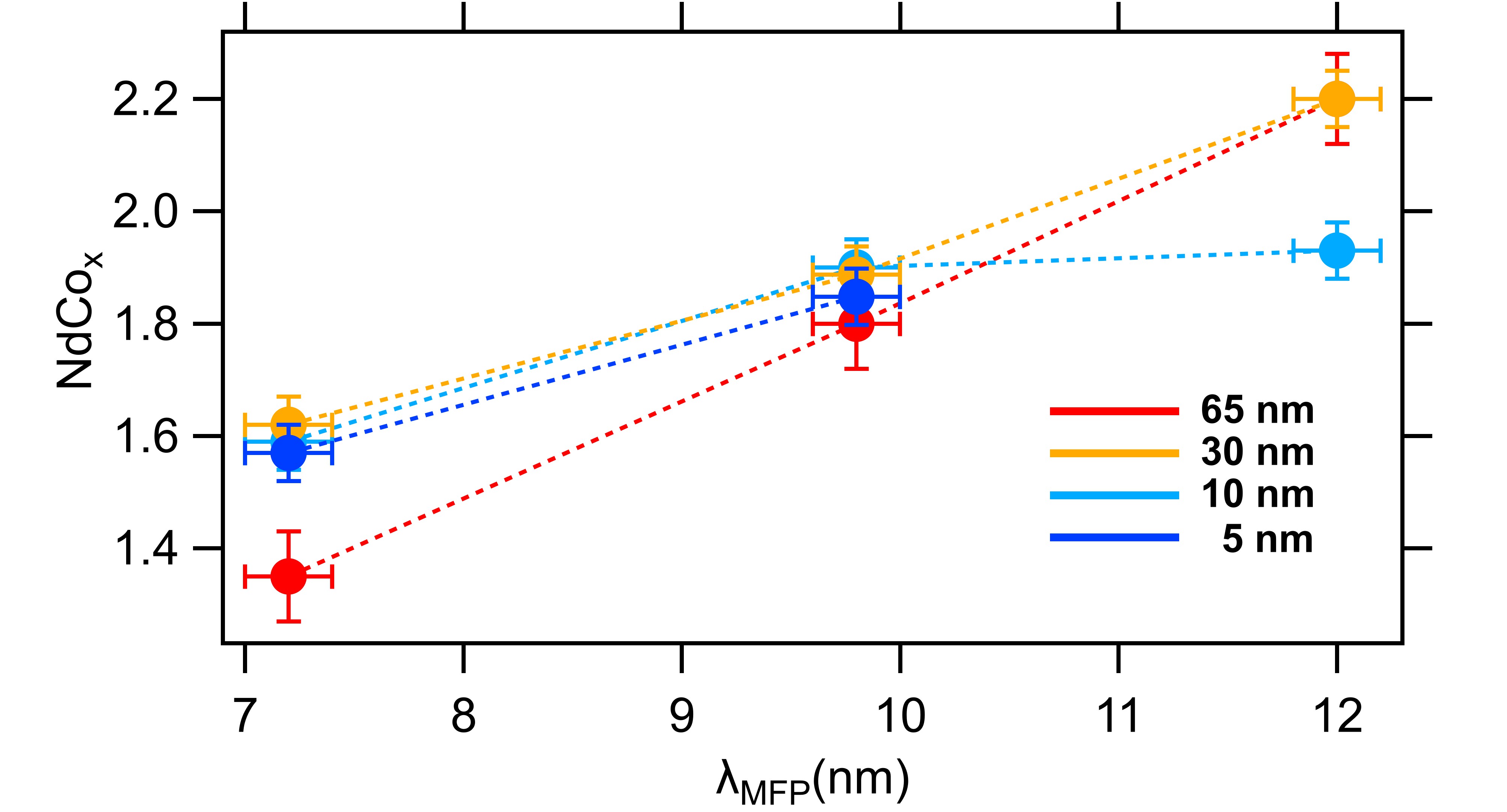}
\caption{Relative atomic concentrations of Co to Nd versus the inelastic mean free path of the photoemited electrons ($\lambda_{IMFP}$) in NdCo$_{4.6}$ films of thickness 5 nm (blue), 10 nm (light blue), 30 nm (orange) and 65 nm (red). \label{NdtoCo_concentration}}
\end{figure}

Figure \ref{NdtoCo_concentration} shows the relative atomic concentrations of Co with respect to Nd obtained in the analyzed samples at three diferent probe depths. The Co-to-Nd atomic concentrations were lower than the nominal values in all the cases. In addition, the Co atomic concentration increases linearly with the related $\lambda_{IMFP}$ of the excited electrons in the 30 nm and 65 nm  thick samples. The concentration saturated in the 10 nm thick sample because its thickness was shorter than the related $\lambda_{MFP}$.  The linear increase in the concentration with the increasing probe depth agrees with a layer of Nd segregated from the alloy whose thickness has to be a fraction of the $\lambda_{MFP}$ used in the experiment. This linear increase was different in the 65 nm film than in the thinner films. This sample has the lowest Co-to-Nd concentration at the lowest $\lambda_{IMFP}$, indicating that the segregated layer was the thickest of all the analyzed samples. The estimated thickness of the Nd top layer ranges between 3 and 4 nm, with a Nd atomic concentration in the top layer of approximately four times the concentration in the compound layer, NdCo$_{4.6}$. This variation in atomic concentration is not far from the obtained by assuming that the top layer is made of Nd$_{2}$O$_{3}$, the most common neodymium oxide, which has 2.5 times a neodymium atomic concentration of NdCo$_{4.6}$. The thicknesses of the layer is in the range of the obtained by XRR and the estimated from the intensity and shape of the photoemission background, as will discussed later. The difference in the slope of the Co-to-Nd atomic concentration versus $\lambda_{IMFP}$ between the samples with 30 nm and 65 nm thicknesses can only be obtained by assuming different values of $\lambda_{IMFP}$ for each of the films, which is coherent with the differences in the density of the films at the top layers. This result qualitatively agrees with the differences observed by XRR for the top layer of the two samples which showed a gradation in intensity in the thinner samples and a better defined layer in the thicker sample.

\begin{figure}
\includegraphics[width=8 cm]{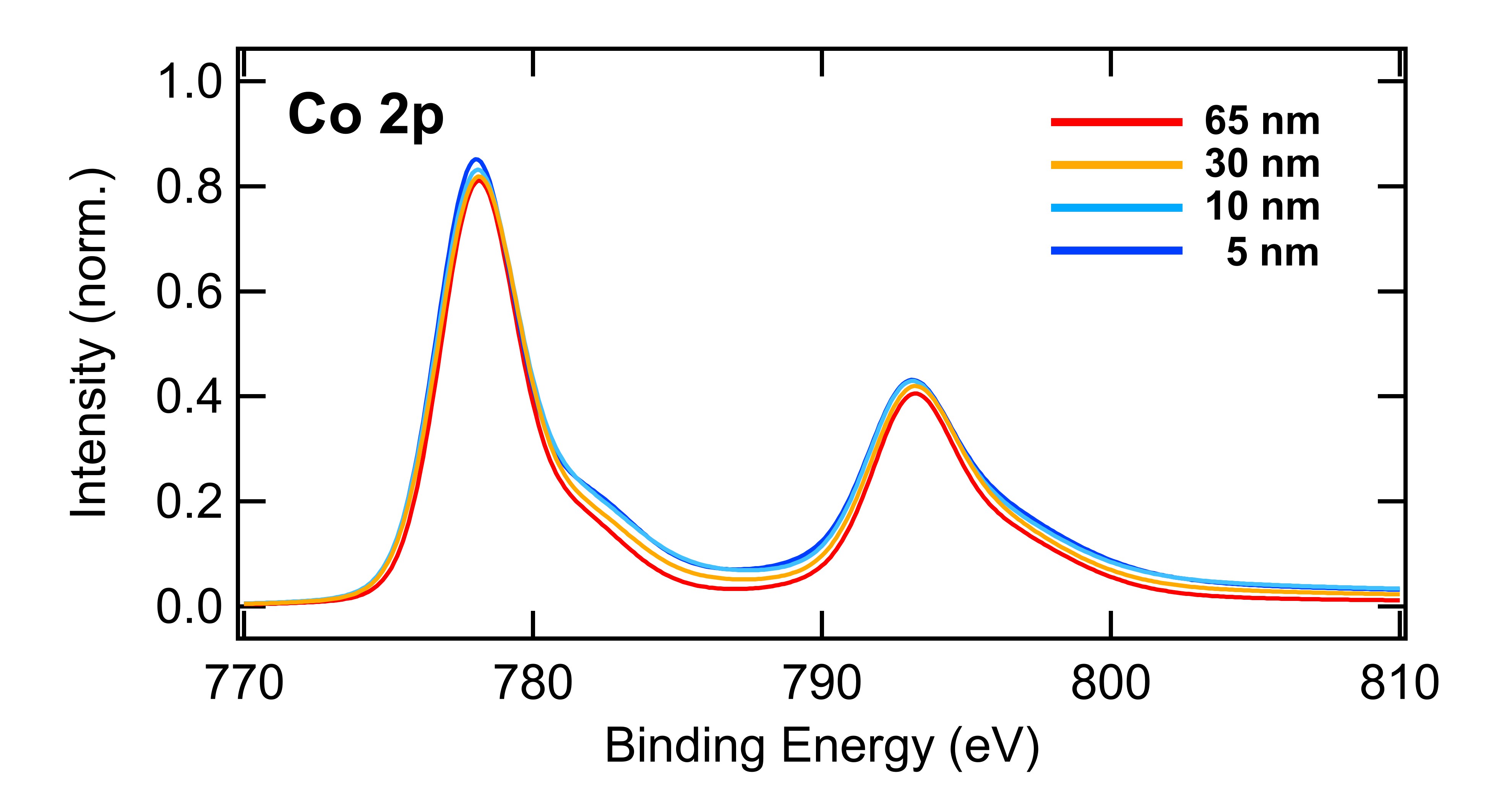}
\caption{Comparison of the fitted Co 2p HAXPES spectra of the analyzed NdCo$_{4.6}$ films obtained at 10 keV photon energy. Thickness: 5 nm (blue), 10 nm (light blue), 30 nm (orange), 65 nm (red).  \label{Fig_Co2p_comp_fit}}
\end{figure}

\begin{figure}
\includegraphics[width=8 cm]{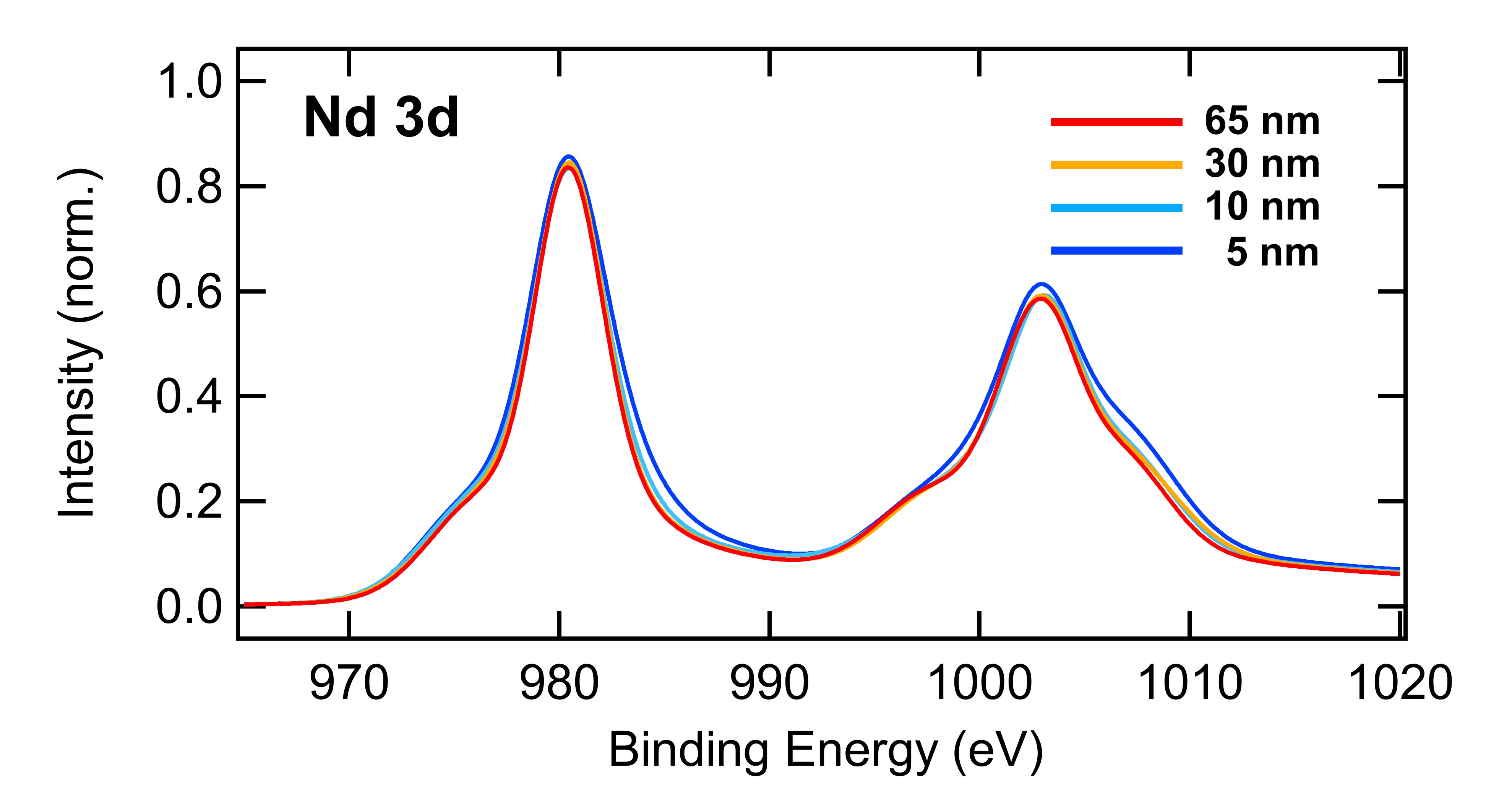}
\caption{Comparison of the fitted Nd 3d HAXPES spectra of the analyzed NdCo$_{4.6}$ films obtained at 10 keV photon energy. Thickness: 5 nm (blue), 10 nm (light blue), 30 nm (orange), 65 nm (red).  \label{Fig_Nd3d_comp_fit}}
\end{figure}

The chemical nature of the alloy in the photoemission-probed region also changes depending on its thickness. Figure \ref{Fig_Co2p_comp_fit} shows a comparison of the fitted Co 2p spectra used for the spectra of the samples measured at 10 keV. No clear differences were observed in the position and features of the peaks. The only visible difference is the width of the peaks, which is wider in the thinnest film and decreases gradually with increasing thickness. This effect was also observed using a photon energy of 7 keV, but the differences in widths between the samples were smaller. The width of the peaks in the thicker sample (65 nm) was the smallest and was equivalent to that of the pure cobalt film. The changes observed in the Nd 3d spectra between the samples, shown in figure \ref{Fig_Nd3d_comp_fit}, have a similar trend to that of cobalt although the thinner sample showed more visible variations in intensity for the components at 1000 eV and 1008 eV BE, respectively. These changes are interpreted as an increasing order in the bonding of cobalt and neodymium with increasing thickness, possibly indicating that the Co to Nd bonding is better defined with increasing thickness. 

\subsubsection{Inelastic photoemission background}
\begin{figure}
\includegraphics[width=8 cm]{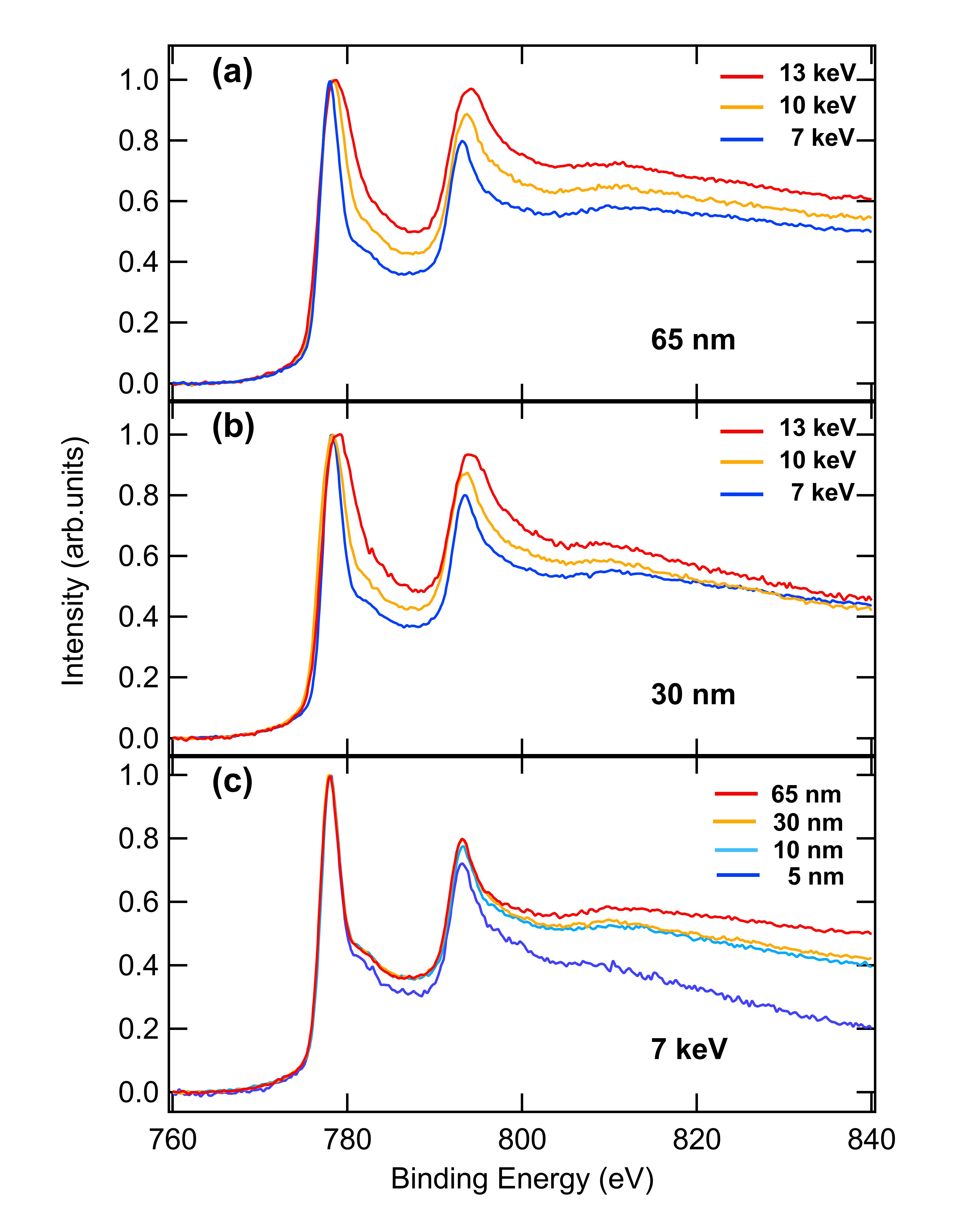}
\caption{Comparison of the Co 2p HAXPES spectra obtained at 7 keV (blue), 10 keV (orange) and 13 keV (red) of NdCo$_{4.6}$ films of thickness (a) 65 nm; (b) 30 nm. (c) Comparison of Co 2p HAXPES spectra obtained at 7 keV of  NdCo$_{4.6}$ films of thickness: 5 nm (blue), 10 nm (light blue), 30 nm (orange), 65 nm (red).  \label{Fig_Co2p_bkg_comp}}
\end{figure}

A detailed study of the background of the Co 2p and Nd 3d peaks is consistent with the presence of a segregated Nd layer in the films. Figure \ref{Fig_Co2p_bkg_comp} compares the Co 2p spectra of the analyzed alloys obtained in the 30 nm and 65 nm thick samples at three different Xray photon energies (panels (a) and (b)), normalized to the intensity of the $2p_{3/2}$ peak. The increase in the width of the peaks with photon energy is due to the decreasing resolution of the  energy of the incident photons and collected electrons. The main changes in the background with increasing photon energy are an increase in its height and a decrease in the negative slope of its tail. Panel (c) compares the background of the analyzed alloys using a photon energy of 7 keV. The background  clearly differs depending on the thickness of the sample. The 5 nm thin sample has the lowest height and the most negative slope of its tail mainly due to its thickness which is smaller than the  $\lambda_{IMFP}$, avoiding electron contribution from deeper regions. The 10 nm and 30 nm thick alloys had approximately the same background indicating a similar configuration of the top region probed by HAXPES for the two thin films. This was unexpected because the probe depth at that energy is larger than 10 nm. This implies differences in the photoemission emitter distribution between the 65 nm and 30 nm thin films because their thickness are higher than the HAXPES probe depth for that photon energy.

\begin{figure}
\includegraphics[width=8 cm]{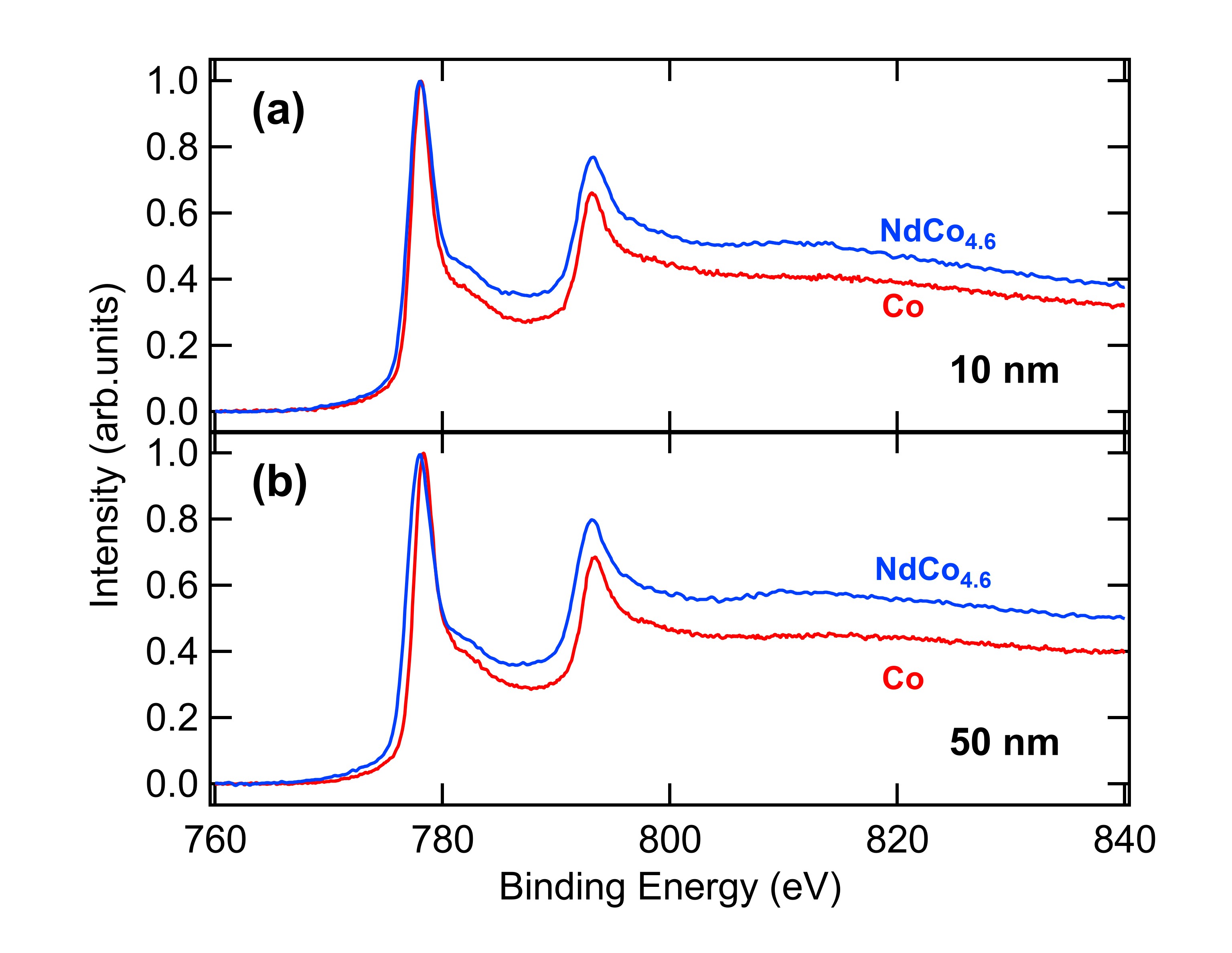}
\caption{Comparison of the Co 2p spectra of NdCo$_{4.6}$ alloy (blue) and pure cobalt (red) taken with 7 keV photon energy for thickness (a) 50 nm (65 nm for the alloy) and (b) 10 nm.\label{fig_Co_vs_NdCo}}
\end{figure}

Figure \ref{fig_Co_vs_NdCo} compares the Co 2p spectra measured at 7 keV photon energy in pure cobalt and in NdCo$_{4.6}$ thin films at two different thickness, 10 nm and 50 nm (65 nm for the NdCo$_{4.6}$ thin film). The spectra were normalized to the intensity of the Co 2p$_{3/2}$ peak showing clear differences in the intensity of the background, which is higher in the NdCo$_{4.6}$ thin films, proving that, in both cases, cobalt is buried under a layer that is thicker in the NdCo$_{4.6}$ films, consistent with the XRR observations and the conclusions obtained from the previous HAXPES analysis.

\begin{figure}
\includegraphics[width=8 cm]{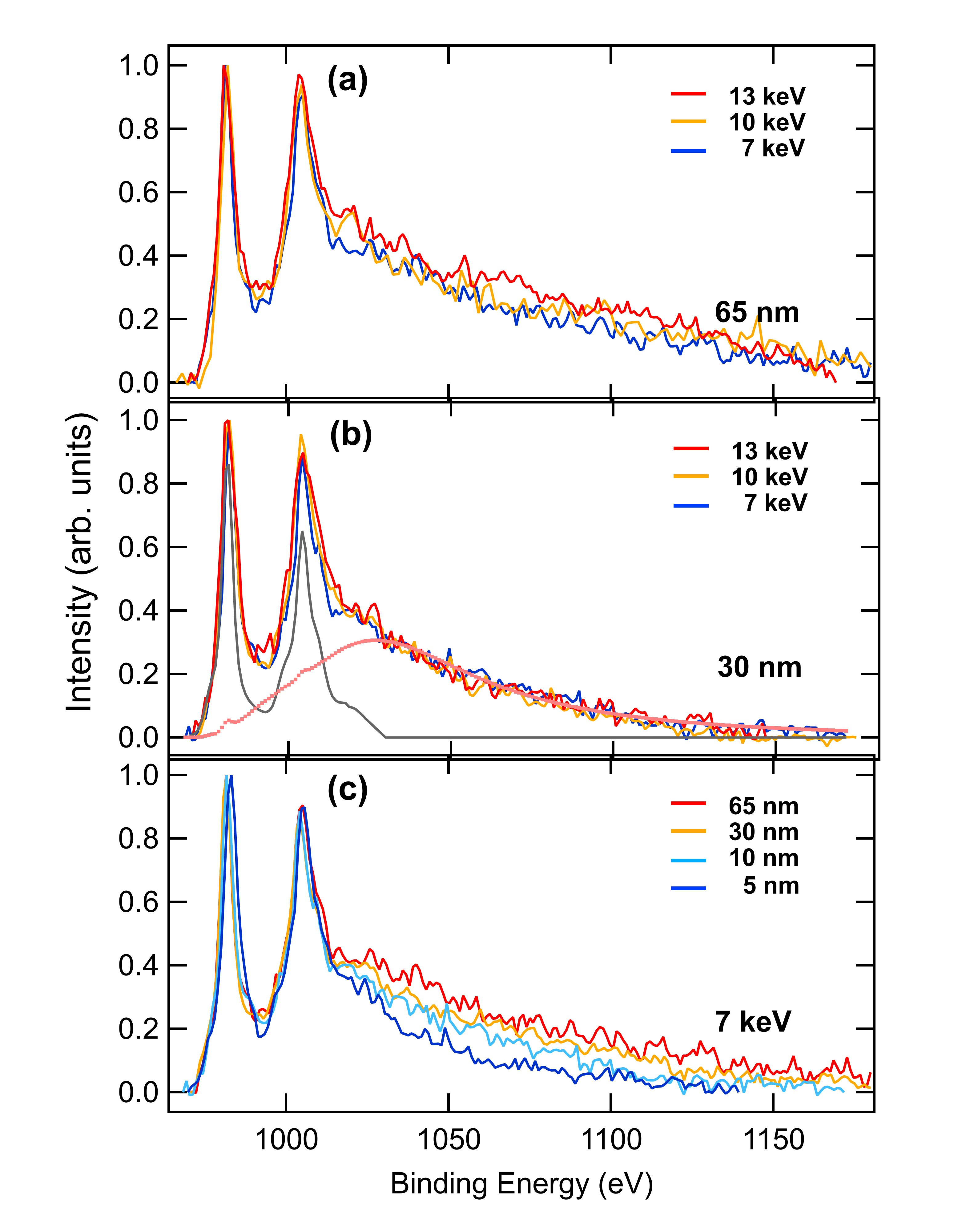}
\caption{Comparison of the Nd 3d HAXPES spectra obtained at 7 keV (blue), 10 keV (orange) and 13 keV (red) of NdCo$_{4.6}$ films of thickness (a) 65 nm; (b) 30 nm. (c) Comparison of Nd 3d HAXPES spectra obtained at 7 keV of  NdCo$_{4.6}$ films of thickness: 5 nm (blue), 10 nm (light blue), 30 nm (orange), 65 nm (red).  \label{Fig_Nd3d_bkg_comp}}
\end{figure}

The determination of the Nd 3d background is not simple because the spectrum is mounted on top of the Co 2s spectrum, which has a lower binding energy (Nd 3d is at 981 eV BE and Co 2s is at 926 eV BE). It was extracted from the Co 2s tail by using a Tougaard-type background for the Co 2s spectra. The consistency of the method was checked by observing that the resulting Co 2s background has a similar evolution  with photon energy and sample thickness as the Co 2p spectra. Figure \ref{Fig_Nd3d_bkg_comp} shows a comparison of Nd 3d backgrounds. The shape of the background in the Nd 3d spectra was different from that observed for Co 2p. The height is smaller and the tail slope was more negative. The background of the thinner alloys does no change when the photon energy ($\lambda_{MFP}$) varies. These changes apparently occurs only in the background height of the 65 nm thick sample, but they are of a much lower magnitude than that observed of the Co 2p background. In fact, the background of Nd 4s, which has the largest tail of all measured Nd excitations, show no changes whit photon excitation in any of the analyzed NdCo$_{4.6}$ films. For a fixed photon energy, the background increases in height with increasing thickness, as occurs in cobalt, but the change in the slope of the background tail is much less pronounced.

The backgrounds of the Co 2p and Nd 3d peaks were simulated using the model described by Tougaard \cite{Tougaard_good} and calculated using the recursive algorithm described in Ref.\cite{Struggling_1979}. The simulation calculates the distribution of the energy of the expected losses for electrons with an initial distribution of kinetic energies, which is the photoemitted spectrum, traveling through a medium with a defined $\lambda_{IMFP}$ a distance $z$. The cross section used for a inelastic scattering event was the called "`universal cross section"', which has been broadly tested in metals.  It only depends on the amount of energy loss, $T$, by the expression: $K(T)=\frac{BT}{\left(C^{2}+T^{2}\right)^{2}}$, where B was set to 3000 eV$^{2}$ and C=1643 eV$^{2}$. The dependence on kinetic energy is through $\lambda_{IMFP}$, which sets the probability of an inelastic scattering event after traveling a distance $z$. The final background was the sum of all contributions from different depths and kinetic energies. The simulation program was a recursive method that was initially fed with the related spectrum without a Shirley-type background and it was recurrently fed with the spectrum substracted from the obtained background in each loop until the best spectrum and background fit were reached. The final simulated spectrum was the sum of the simulated background and elastically scattered spectrum whose intensity decreased according to the expected inelastic loss events defined by $\lambda_{IMFP}$.

\begin{figure}
\includegraphics[width=8 cm]{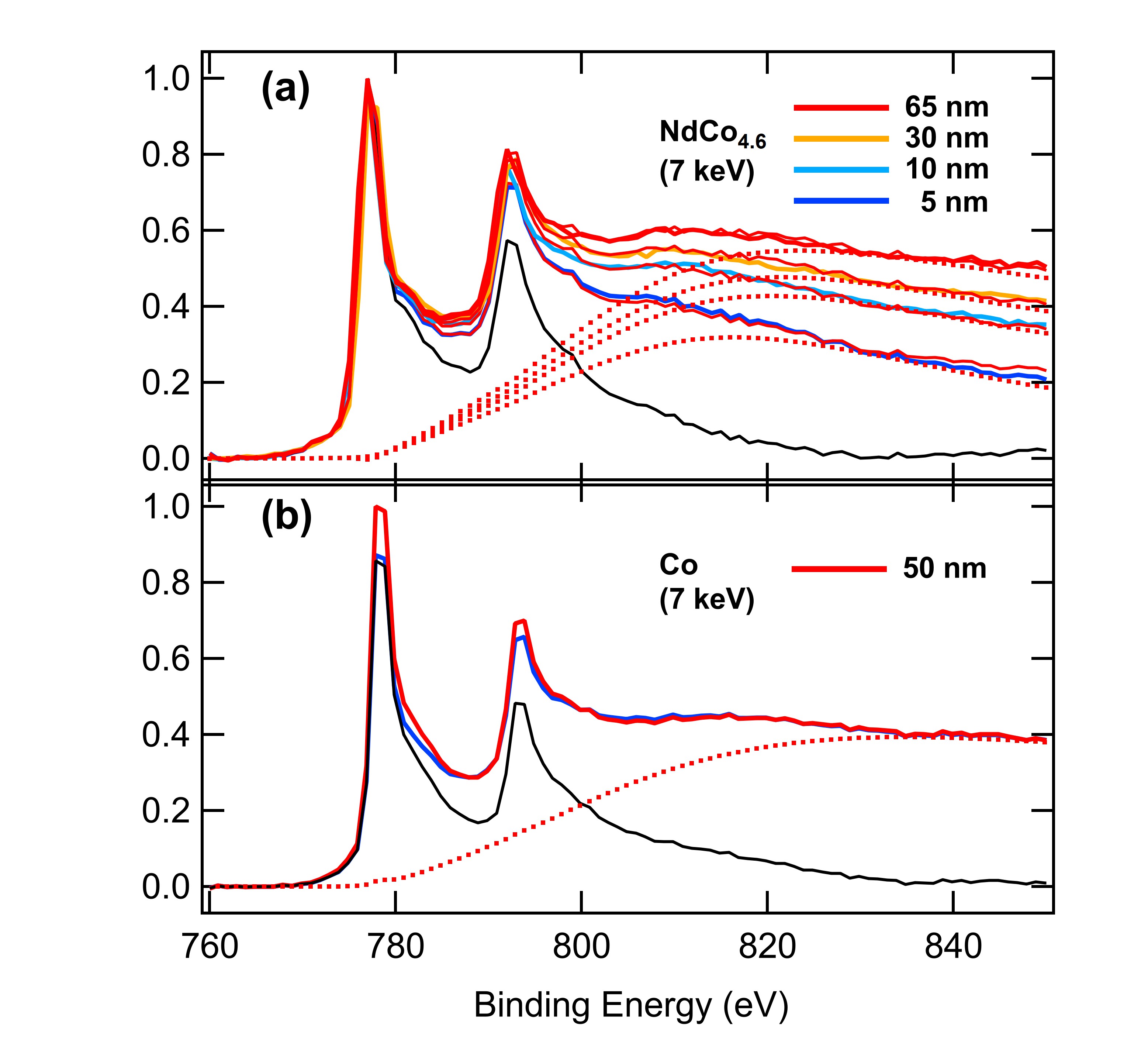}
\caption{Comparison of the Tougaard type fitted backgrounds for the Co 2p HAXPES spectra obtained at 7 keV of (a) NdCo$_{4.6}$ films of thickness: 5 nm (blue), 10 nm (light blue), 30 nm (orange), 65 nm (red), and (b) pure cobalt 50 nm thick film.  \label{Fig_Co2p_Tougaard}}
\end{figure}

Figure \ref{Fig_Co2p_Tougaard} shows the obtained backgrounds for the Co 2p spectrum of all the analyzed alloys and the thicker pure cobalt film using a photon energy of 7 keV. An example of the background fitting of the Nd 3d spectra is shown in figure \ref{Fig_Nd3d_bkg_comp}. The distribution of the photoemission emitters within the films used for the fits consists of a step function whose parameters are the length of the step, related to the extension in depth within the film from where secondary electrons contribute to the background, and the position with respect to the surface of the sample, that is, the buried depth of the photoemission emitters. The fitted curves used $\lambda_{IMFP}$ values corresponding to pure cobalt. The height of the background is more sensitive to the thickness of the buried layer, whereas the slope of the background tail had a higher dependence on the emission electron depth, where more inelastic scattering events are expected to convey a stronger broadening of the energy of the scattered electrons. The minimum variation in thickness was 1 nm. This simple model allowed consistent results compared with more complex photoemission emitter distributions with more parameters to fit, showing the limitations of the technique for finding details in this distribution.   

\begin{figure}
\includegraphics[width=8 cm]{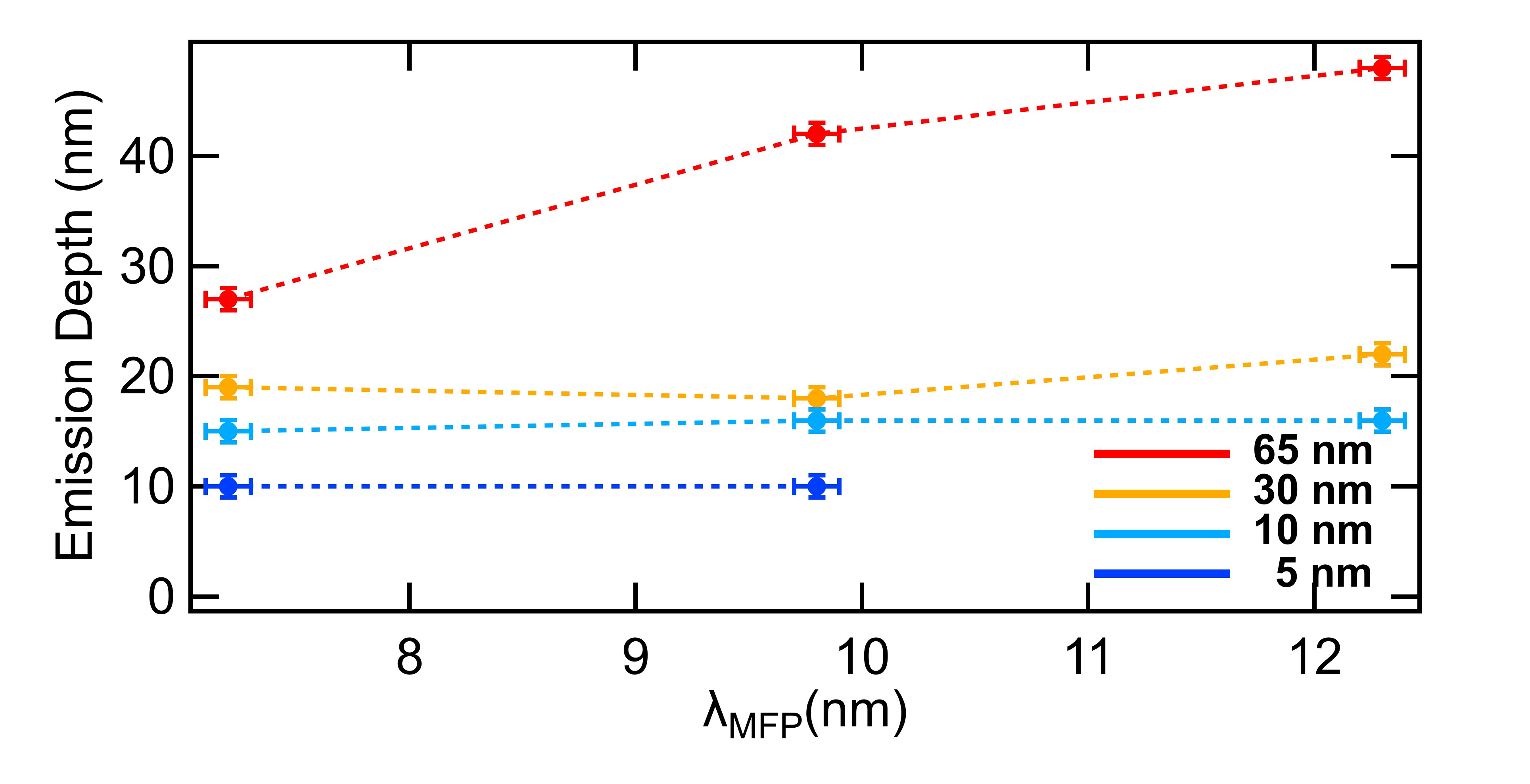}
\caption{Lengths from where secondary electrons contributed to the background of the Co 2p spectra in NdCo$_{4.6}$ films of thickness: 5 nm (blue), 10 nm (light blue), 30 nm (orange), 65 nm (red)  \label{depth_Co2p_Tougaard}}
\end{figure}

The background of the pure cobalt films was used as a reference for the parameters used in the fits. The emission depths obtained for the 10 nm and 50 nm thick Co films, assuming no burying layer, were almost twice as larger as expected, yielding 18 nm and 32 nm, respectively, for 7 keV photons ($\lambda_{IMFP}=$7.2 nm). In both cases, the intensity of the elastically scattered photoemision peaks must be reduced from the expected value. Similar behavior was observed in the NdCo$_{4.6}$ films, indicating that the simulated backgrounds were smaller than the measured values. The applied reduction in the intensity of the photoemission peaks with respect to the background was not the same for all the samples and was higher in the thinner films. The emission depths obtained for the NdCo$_{4.6}$ films at different energies are shown in figure \ref{depth_Co2p_Tougaard}. There were clear differences between the thinner and the thickest samples. The ratio between the emission depths of the 30 nm and 65 nm films is similar to their thickness ratio. However, the same ratio was smaller for thinner films. The model gives the right intensity of the photoemitted for the thickest film but overestimates the intensity in the thinner films. One possible reason for this misfit is cobalt diffusion into the substrate, whose effect will be stronger when the film is thinner. Both, pure cobalt and NdCo$_{4.6}$ compound films have significantly lower magnetization values in the thinner films, as shown in the previous section, whose origin should be cobalt diffusion into the substrate. From figure \ref{depth_Co2p_Tougaard}, the 5 nm and 10 nm films gets the expected emission depth by dividing their experimental one by the same factor than that for the 10 nm cobalt film. However, to have the same intensity in the photoemission peak, all of the films require a burying layer of about 2$\pm1$ nm thickness assuming a $\lambda_{IMFP}$ of cobalt. This result can be estimated directly from the comparison between the pure cobalt and NdCo$_{4.6}$ films shown in figure \ref{fig_Co_vs_NdCo} by noting that the difference in the background intensity is due to a thicker burying layer in NdCo$_{4.6}$ films. If the difference in thickness is $\Delta z$, then $\Delta z\approx \lambda_{IMFP}\ln\left(I_{NdCo_{4.6}}/I_{Co}\right)$, yielding 2 nm for the 10 nm film and 2,5 nm for the 50 nm film, using a $\lambda_{IMFP}$ corresponding to an electron density similar to the obtained by XRR. This result is consistent with that of the XRR analysis. It is also similar to the estimated in DyCo \cite{DyCo_XMCD} signaling that the effect can be extrapolated to other RE-TM compounds.

The simulations of the background for the Nd 3d spectra do not admit a burying layer, which is consistent with the presence of a segregated layer of Nd on top of the alloy films. However, this does not explain the lack of the background shape dependence on  $\lambda_{IMFP}$. It also underestimated the height of the background with respect to the elastically scattered part of the spectra as it occurred in the background of the Co 2p spectra. These failures in the description of the background are independent of the chosen emitter distribution. We suggest that this misbehavior in the simulations might be related to the failure of the model to explain the background for the dilute concentration of Nd within the alloy at depths above the segregation layer. This could be a different situation than that of the photoemitted electrons from Cobalt which are majority within the alloy. 

The low density of Nd atoms would make less probable to detect photoemitted electrons which have been multiple inelastically scattered, giving more weight to single inelastical scattering events. This explains the constant background shape under $\lambda_{IMFP}$ variations and the underestimation of the background height. In this case, the Nd 3d background differences found between the samples are mainly related to their sample thickness, consistent with the presence of neodymium emitters within the films at regions deeper than the segregated layer. 

This coarse sensitivity of the background shape to determine the structure of the segregated layer is expected because of the probe depth which in this case is of the order of three to four times $\lambda_{MFP}$, whereas the main segregated layer is just a fraction of $\lambda_{MFP}$. Thus, it was not possible to relate the observed differences in the Nd 3d backgrounds to the actual differences in the distribution  of neodymium between the samples. Such a description can be performed using photoemission spectroscopy with a shorter probe depth.



\section{CONCLUSIONS} 

Nd segregation at the surface of NdCo thin films was investigated using XRR and HAXPES in samples exhibiting perpendicular magnetic anisotropy (PMA) above a critical thickness of 30–40 nm. RE segregation was detected across all thicknesses studied, ranging from 5 nm to 65 nm. XRR revealed changes in the top layer precisely at the thickness where the compound began to show an increase in PMA energy.

The Co-to-Nd concentration ratio, derived from the Co 2p and Nd 3d spectra, was consistent with the XRR observations, indicating an enhanced Nd concentration in the top layer of the thickest films. Analysis of the Co 2p and Nd 3d spectral backgrounds further supported the presence of a segregated Nd layer. The thickness of this segregated layer was estimated to be approximately 2–3 nm, based on measured atomic concentrations and photoemission background simulations employing step-function distributions of electron emitters.

The observed Nd segregation at different stages of growth proofs the diffusion of Nd atoms towards the surface of the film during thin film deposition. This fact together with the large compressive strain previously probed by EXAFS and the progressive increase in PMA energy with thickness, suggests that the origin of PMA in these films is linked to the minimization of strain energy during growth. Specifically, the larger atomic volume of Nd compared to Co imposes strain when Nd incorporates into the Co sublattice. Energy minimization favors Nd atomic environments elongated along the vertical direction, thereby relieving strain and promoting the perpendicular anisotropy observed in these films.

\begin{acknowledgments}
We acknowledge the European Synchrotron ESRF, the Spanish Ministerio de Ciencia, Innovacion y Universidades, and the Consejo Superior de Investigaciones Científicas for provision of synchrotron radiation at BM25 and for financial support through the projects PIE 2010-6-0E-013, 2021-60-E-030 and CEX2024-001445-S. J. D. and J. R-F. acknowledges the Spanish Ministerio de Ciencia, Innovacion y Universidades support under grants 104604RB/AEI/10.13039/501100011033 and PID2022-136784NB and by Agencia SEKUENS (Asturias) under grant UONANO IDE/2024/000678 with the support of FEDER funds.

\end{acknowledgments}

\section*{Data Access Statement}
Research data supporting this publication are available from the Universidad de Oviedo repository located at https://hdl.handle.net/10651/80843.
\section*{Conflict of Interest Declaration} 
The authors declare that they have no affiliations with or involvement in any organization or entity with any financial interest in the subject matter or materials discussed in this manuscript.

\end{document}